# A robust methodology for inferring physiology of a protein family: application to $K^+$-ion channel family

Ashok Palaniappan

We are interested in the subtle variations of function among the members of a protein family. A protein family is usually subdivided into subfamilies based on functional differences. Existence of this functional diversity is essential for the successful performance of physiological roles expected of the family. This presents a unique problem: there must be preservation of the active site; simultaneously there should be specificity of protein action according to subfamily. Though the classification into subfamilies is by no means a formalized one, it is most times based on the character of regulation of the primary function. The function of a subfamily is a modification of when the protein performs its function, for example, by changing the protein's sensitivity to regulatory factors. Rarely, a subfamily possesses a function completely different for its family. A study of these details is necessary for understanding the fine-tuning of protein function. I describe a theory for studying subfamily-based functional specificity and then validate it with an example application to deciphering the residue-level basis of fine functional variations in the diverse set of $K^+$-channel subfamilies. I provide specific results that will be useful to channel physiologists, whereas the strategy developed will be widely applicable to problems in comparative and functional genomics.

## I  METHODOLOGY

Protein subfamilies are represented in a genome by multiple members, which are almost identical in sequence, but nevertheless conservatively accumulate mutations. A differentiated subfamily is the outcome of functional divergence in duplicated copies of the parent gene. When a duplicated gene acquires enough mutations to functionally diverge from the original gene, it becomes a subfamily. Thus it is an ortholog turned paralog. An extant protein family by way of multiple previous duplications and subsequent functional divergence could include many subfamilies. This general mechanism is one of the most important used by evolution for generating new functions. Different subfamilies consist of distinct residue signatures underlying individual biochemical events, and these signatures impart the physiological character of a specific subfamily.

To compare the sequence bases of subtle functional variations among the subfamilies of a given family, it is useful to invent a single sequence that is representative of the subfamily. Such a sequence must preserve invariant residues of a given subfamily, and carefully define appropriate residues for other positions, thus encapsulating the information about a subfamily in a concise form. In this manner, we obtain a manageable number of sequences representing the subfamilies of a protein family. Comparisons of corresponding multiply-aligned positions in such consensus sequences can reveal the subfamily-level basis of functional specificity. In particular, the following types of information emerge:

1. positions conserved in all subfamilies are precisely those that are important for the function of the protein family.
2. positions differentially conserved, that is conserved in some subfamilies and variable in others, suggest specific functions important to certain subfamilies and absent in others.
3. variable positions are more difficult to analyze; in some instances, they could be residues that are largely inconsequential to protein function, in others they could signify residues on exposed loops that are important for function, thus implying a hotspot for conferring functional variability within a protein family.

When these positions are analyzed in combination with evidence from structural and physiological sources, it becomes possible to deduce from sequence alone the residue-level basis of functional specificity. These principles constitute the basis of our theory for studying and inferring physiological characteristics of a protein family from sequence information.

In order to apply the theory, consensus sequences must be constructed first. Several strategies have been proposed to derive consensus sequences. The simplest is to compute the consensus residue as the most frequently occurring residue for a position. This calculation is simplistic because:
1. conservative substitutions are not taken into account;
2. position-specific weighting is ignored; and
3. pseudocounts are not included.

The technique described in (Henikoff and Henikoff 1997) was among the best performing for the construction of consensus sequences for detecting homologs. A position-dependent pseudo-count was used in their method, and consensus residues are computed only for the conserved regions of an alignment, while the residues of an individual sequence are substituted in the variable portions. Thus the technique 'embeds' consensus residues into a sequence member of the family. This overcomes two problems with other strategies for computing consensus sequences:
1. global strategies compute consensus residues over all positions of the alignment. This compromises performance by ignoring uncertainties in alignment.
2. motif-based strategies compute consensus residues only over significantly conserved regions. The variable positions are ignored in the consensus sequence.

This scores an effective compromise between global and motif-based strategies.

## II    APPLICATION: INFERRING PHYSIOLOGY OF $K^+$-CHANNELS

Below we implement the theory to the understanding of functional diversity in the $K^+$-ion channel family. In addition to providing a concrete basis for validating the theory, this specific example leads to powerful insights about channel function. The comprehensive identification of all $K^+$-channels in the human genome is described in (Palaniappan and Jakobsson 2006). The representation of each subfamily in the human genome is quantified in the table shown below.

| № | Subfamily | Description | Genes |
|---|---|---|---|
| 1 | KCNA | voltage-gated channel subfamily A (Shaker-related subunits) | 11 |
| 2 | KCNB | voltage-gated channel subfamily B (Shab-related subunits) | 3 |
| 3 | KCNC | voltage-gated channel subfamily C (Shaw-related subunits) | 7 |
| 4 | KCND | voltage-gated channel subfamily D (Shal-related subunits) | 4 |
| 5 | KCNF | voltage-gated channel subfamily F ('silent' modifiers) | 1 |
| 6 | KCNG | voltage-gated channel subfamily G ('silent' modifiers) | 4 |
| 7 | KCNS | voltage-gated channel subfamily S ('modifier' subunits) | 5 |
| 8 | CNGA | Cyclic nucleotide-activated channel α subunit (CNG α) | 5 |
| 9 | CNGB | CNGA-regulatory cyclic nucleotide activated β subunit (CNG β) | 2 |
| 10 | KCNH | human ether-a-go-go related channel (Herg) | 12 |
| 11 | KCNQ | KQT-like voltage-gated channel (KvLQT) | 11 |
| 12 | KCNK.1x2 | First pore of two-pore channels | 23 |
| 13 | KCNK.2x2 | Second pore of two-pore channels | 24 |
| 14 | HCN | Hyperpolarization-activated cyclic nucleotide-gated channels ($I_h$ channels) | 5 |
| 15 | KCNN | Intermediate/small conductance calcium-activated channel (IK/SK) | 5 |
| 16 | KCNJ | Inward-rectifying channel (IRK) | 17 |
| 17 | KCNMA | Large conductance calcium-activated channel (BK) | 4 |
| 18 | KCNSlack | Sequence like a calcium-activated $K^+$-channel (Slack) | 3 |

Table 1. Representation of $K^+$-channels according to subfamily in human genome. The rationale for independently considering the two pores of the two-pore channel subfamily, and other such details, are described in (Palaniappan 2005).

Each subfamily's consensus sequence was derived and used to construct a multiple alignment of all subfamilies. Fig.1 represents the alignment of consensus sequences of the 18 human $K^+$-channel subfamilies. An analysis of notable features from the alignment revealed the following:

1. Selectivity filter motif :

    GYG is the predominant selectivity filter triplet. Three subfamilies have the GFG sequence. These are the hERG, first pore of the two-pore channels, and large-conductance $Ca^{2+}$-activated channels (KCNMA). The presence of the GFG filter in KCNMA contradicts the result from an earlier study that weakly conducting channels appear to have a greater preponderance of the GFG filter. Both CNG channels, i.e., the αs and the βs have unique filter properties that relax their ion selectivity. The primary structure basis for these properties can be studied from the alignment above. The aromatic residue (Tyr or Phe) in the middle of the canonical $K^+$ selectivity filter is absent in the case of the CNG channels. In fact, there is a gap in the alignment in this position for the CNG channels, indicating that there was a specific deletion mutation of the codon for the middle residue of the canonical $K^+$ selectivity filter. It is likely that this evolutionary event relaxed the selectivity of an ancestral $K^+$ channel, and made the pore tolerant to the passage of other cations too. The selectivity filter glycines remain conserved. In addition, the CNG channels possess an extra Gly or Pro within three residues of the selectivity filter; presumably these residues foster the flexibility necessary for a less discriminating selectivity filter. This inference provides a basis for the reconstruction of the cation selection of the CNG channel at an atomic level from sequence homology to $K^+$ channels.

2. P-x-P voltage-gating motif:

    Eukaryotic voltage-gated ion channels have been observed to possess a P-x-P motif, where x is Val or Ile, in the outer S5 helix (Long and others 2005). It is located two residues intracellular to the bundle crossing. We see that it is found only in the voltage-gated $K^+$-channel subfamilies, namely Shaker, Shal, Shaw, and Shab. In channels that are somewhat voltage-sensitive, one of the prolines is present, and in electrically-silent channel subfamilies, the motif is absent.

```
kcna      LGQTLKASMRELGLLIFFLFIGVILFSSAVYYFAE-IPDAFWWAVVTMTTVGYGDMHP-P
kcnc      LGHTLRASTNEFLLLIIFLALGVLIFATMIYYYAERIPIGFWWAVVTMTTLGYGDMYP-P
kcnb      LGFTLRRSYNELGLLILFLAMGIMIFSSLVFFFAE-IPASFWWATITMTTVGYGDIYP-P
kcnd      LGYTLKSCASELGFLLFSLTMAIIIFATVMFFYAE-IPAAFWYTIVTMTTLGYGDMVP-P
kcnf      LTYALKRSFKELGLLLMYLAVGIFVFSALGYYTME-IPQSFWWAIITMTTVGYGDIYP-P
kcng      LGLTMRRCYREFGMLLLFLCVAMALFSPLVYYVAE-IPACYWWAIISMTTVGYGDMVP-P
kcns      LGFTLRHCYHEVGCLLLFLSVGISIFSVVAYYSVE-IPCCWWWATVSMTTVGYGDMHP-P
cnga      NYPNIFRISNLMMYIFIIIHWNACIYYAISKKYIGFYIYCFYWSTLTLTTIG-GETPP-P
cngb      DKAYIYRVIRTTGYLLYILHINACVYYWASNAYQGIYIRCYYWAVRTLITIG-GGLPD-P
herg      SEYGAAVLFLLMCVFALIAHWMACIWYAIGNMVERPYVTALYFTMSSLTSVGFGNVSP-P
kcnq      ----VYAHSKELITAWYIGFLCLIFSSFFVYYLAEKYADALWWGLITLTTIGYGDKYP-P
kcnk.1x2  ---IVHAVLLGFGCLCFLCIPAVVFSHVLED-----FFEAFYFCFITLTTIGFGDYVAGQ
kcnk.2x2  GGLWWWRFLCLFCVYVCYLVLGAVVFWALEGGPHECFPSAFFFACTVITTIGYGHMAP-P
hcn       LASAVMRICNLIGMMLLLCHWDGCLQFLVPMLLQD-YSYALFKAMSHMLCIGYGRQAP-P
kcnn      VMKTYMTICPGTVLLVFSISLWIIAAWTVVRVCER-FLGAMWLISITFLSIGYGDMVP-P
irk       MKWRWMMLIFSMAFLCSWLFFGMIWWLIAYIAHGDLFTSAFLFSIETQTTIGYGFRCVVE
kcnma     -TSNSVKFSKLLSIFLSTWFTAAGFIHLVENSGDPWYWECIYLVMATMSTVGFGDVYA-A
kcnslack  ILRTQSAMFNQVLILFCTLLCLVVFTGTCGIQHLERLFTSFYFCIVTFSTVGYGDVTP-P
consensus l---l------lgllil--l-igvilf--m-----e-y--afwwavvtmttvGyGdm-p-p

kcna      VTVGGKIVGSLCAIAGVLTIALPVPVIVSNFNYFYHH
kcnc      QTWSGMLVGALCALAGVLTIAMPVPVIVNNFGMYYSS
kcnb      KTLLGKIVGGLCCIAGVLVIALPIPIIVNNFSEFYKK
kcnd      KTIAGKIFGSICSLSGVLVIALPVPVIVSNFSRIYHH
kcnf      KTTLGKLNAAISFLCGVIAIALPIHPIINNFVRYYNN
kcng      RSVPGQMVALSCILSGILLMAFPITFIFHTFSHCYHL
kcns      DTHAGKFFAFMCILCGILVVAMPITIIFNKFSHCYRQ
cnga      VRDEEYLFMVVDFLIGVMIFATIMGNMGSMIYNMM--
cngb      QTLFEIVFQLLNYFTGVFVFSVMIGQMRDVIGAAA--
herg      NTDSEKIFSICVMMIGSLMYATIFGNVTAIIQRMYSS
kcnq      QTWNGRLIAACFTLFGISFFALPAGILGSGFALKVQQ
kcnk.1x2  KFPWYKPFVWCWILLGLAWMALFLNWTFCELHELKKF
kcnk.2x2  LTDGGKAFCMFYALFGIPPASLALVATLRHCLLPVLS
hcn       VSMTDIWLTMLSMIVGATCYAMFIGHATALIQSLL--
kcnn      HTYCGKGVCLCTGIMGACCTALVVAVVARKLEFTKAE
irk       ECPEAIFMLIFQSILGCIINAFMCGCMFAKMARPKKR
kcnma     KTSLGRTFMMFFTLGGLIMFANYIPEMVELFGNKRKY
kcnslack  KIWPSQLFVVIMICVALVVLPIQFYYLVYLWWMERQK
consensus -t--gkif-ll--l-gvlvial-v--i---f------
```

Figure 1. Multiple alignment of the consensus sequences of K$^+$ channel subfamilies. This is the essential step in comparative analysis of physiological variation found in the subfamilies within a protein family.

3. Gating hinge:

   The gating hinge is Gly in all subfamilies except the Slack subfamily which have Ala instead. The presence of a small residue at this position allows the channel to change conformation during gating by minimising strain in the peptide backbone.

4. T-x1-TT-x2 motif:

   The sequence immediately preceding the selectivity filter has been designated the T-x1-TT-x2 motif, where x1 and x2 denote variable positions (Heginbotham and others 1994). The complete alignment enables a new characterization of x2.

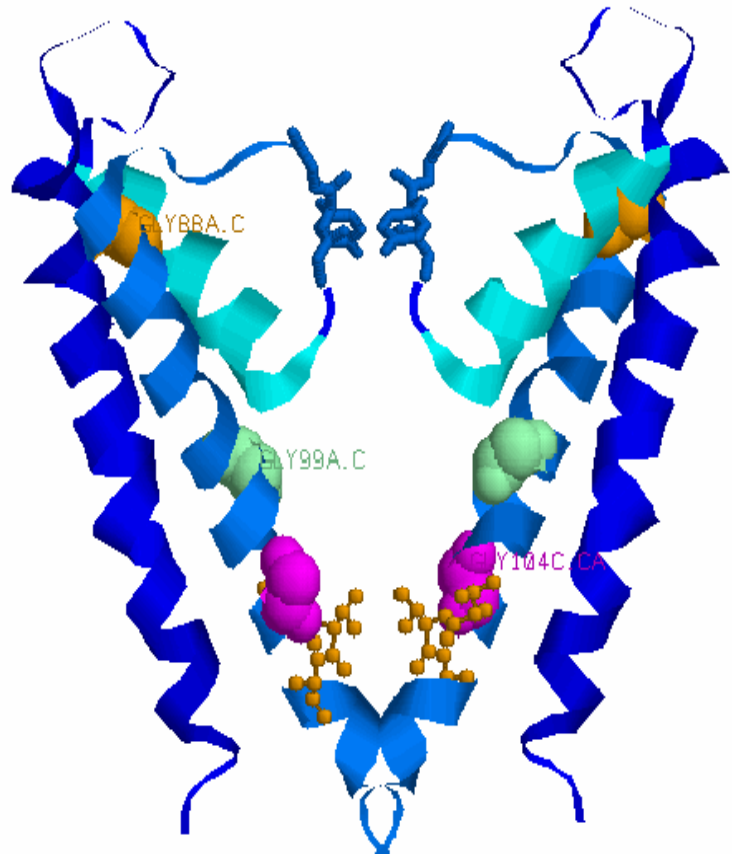

Figure 2. A cartoon representation of the pore structure of bacterial KcsA K$^+$-ion channel (Doyle and others 1998). Only two opposing monomers are shown for clarity. The structure is expected to be conserved in all K$^+$-channels including human K$^+$-channels; see (MacKinnon and others 1998), (Palaniappan and Jakobsson 2006). Notable features discussed in text are highlighted as follows: selectivity filter in blue wireframe, PxP motif in orange ball and stick, Gly gating hinge in the middle of S6 in green spacefill, intersubunit bundle crossing point in magenta spacefill and another Gly at the N-term of S6 in orange spacefill models.

x2 is always a Leu, Ile, or Val, i.e., one of the large hydrophobic amino acids. HCN channels possess a different signature in this region, viz., H-x1-LC-x. hERGs have Ser replacing two of the Thr, while KCNMA, Slack, KCNG, and KCNS have Ser replacing one of the Thr. The second pore of two pore channels show a Val replacing one of the Thr, while the KCNN channels have a Leu in place of one of the Thr. Though the significance of these particular facts is not clear in the context of evolution, they are no doubt very important in guiding the design of experiments.

5. Aromatic cap of pore helix:

   In all voltage-gated $K^+$ channels, there is a triplet of aromatic residues at the N-terminus of the pore helix that function to orient the channel conduction pathway, and providing the necessary rigidity of the structure for strict ion selection. Aromatic residues are known to stably anchor transmembrane elements (Killian and von Heijne 2000). The large-conductance $Ca^{2+}$-activated channels have only one aromatic residue at this position, presumably enhancing the flexibility of the structure for higher ion throughput. The HCN and KCNN subfamilies also have only one aromatic residue at the membrane-water interface of the pore helix; hERG and KCNQ have two.

6. Extended selectivity filter motif:

   Analysis of the extended selectivity filter motif reveals a Pro doublet (P-P) in this region in 14 out of the 18 subfamilies. In contrast to the aromatic cap of the pore helix, this region must be involved in enhancing flexibility necessary in conformational changes. Of the four exceptions, CNG β conserves one of the Pro doublet, KCNK first pore has a Gly, IRK has a Val doublet, and KCNMA has a Ala doublet.

7. Intersubunit bundle crossing point:

   All, except three, subfamilies possess the secondary gating hinge or the intracellular intersubunit bundle crossing, which is an Ala five residues C-term of the gating hinge. CNG β and KCNK second pore subunits possess Ser at this position and Slack has Pro. These statistics are in themselves useful even when the contribution to channel physiology is not evident.

8. Gly 11 residues N-term to gating hinge:

   A last observation concerns Gly that is 11 residues N-term of the gating hinge and that has been observed to be conserved in the $K^+$-channels of many species (Shealy and others 2003). Though the role of this residue is not experimentally characterized, we note that all the voltage-gated $K^+$ channels, SK and BK conserve this position, while subfamilies with a cyclic nucleotide-binding domain

(i.e., hERG, HCN, CNG-α and the CNG-β) have acidic character (Asp or Glu) at this position. Perhaps these differences suggest a deeper physiological significance for this position.

Proceeding from the above account, it would be possible to confidently reconstruct a composite picture of permeation, gating and other salient properties of each $K^+$-channel subfamily. A generalized protocol for the comparative analysis of any protein family includes the following stages:

1. construction of comprehensive database of protein family in genome
2. construction of consensus sequence of each subfamily, followed finally by multiple alignment of all consensus sequences
3. comparative genomics in parallel with study of structural and physiological evidence.

The example we have chosen for demonstrating our theory, the $K^+$-channel family, clearly validates the effectiveness of the theory for comparative genomics at the subfamily-level. There have been many significant efforts in this direction, notably (Lichtarge and others 1996), (Sjolander 1998), (Brown and Sjolander 2006), and (Hannenhalli and Russell 2000). Our methodology appears to be a robust alternative to existing methods, and will usefully supplement them. Moreover it is very simple that it may be considered first in challenging problems of comparative and functional genomics.